\documentclass[letter]{jpsj2}
\usepackage{graphicx}

\def\runauthor{Masato \textsc{Kishi} and Yasuhiro \textsc{Hatsugai}}

\title{%
Dielectric Response of the Interacting 1D Spinless Fermions with Disorder
}
\author{%
Masato \textsc{Kishi}\thanks{E-mail: kishi@pothos.t.u-tokyo.ac.jp} and
Yasuhiro \textsc{Hatsugai}
}

\inst{%
Department of Applied Physics, University of Tokyo, 7-3-1 Hongo\\
Bunkyo-ku, Tokyo 113-8656, Japan \\
}

\recdate{\today}

\abst
{
Dielectric responces of the one-dimentional electron system is
investigated numerically.
We treat an interacting one-dimentional spinless fermion model with
disorder by using the Density Matrix Renormalization Group(DMRG) 
method which is extended for nonuniform systems.
We apply an electric field $E$ to the system 
and calculate dielectric responces.
Dielectric responce of the Mott insulator and the Anderson insulator
are calculated respectively.
Steplike behaviors in $P-E$ curve are obtained which corresponds to
breakdown of the insulating behavior.
For the Mott insulator, the steps originate from generation of kink-pairs.
For the Anderson insulator on the other hand,  
the origin of the steps is a crossing of the localized one particle
energy levels. 
We also treat random systems with interaction.
From one parameter scaling analysis of the susceptibility $\chi$, 
the metal-insulator transition in attractively interacting region 
is confirmed and a phase
diagram of the random spinless fermion model is obtained.
}
\kword{%
Anderson localization, polarization, dielectric response, density
matrix renormalization group
}

\begin{document}
\sloppy
\maketitle

\section{Introduction}
Effects of randomness and interaction in electronic systems are two
major problems in condensed matter physics.
These two problems have
been studied intensively for several decades and a lot of fundamental
results are accumulated.
The presence of disorder entails a localization of electronic states
due to a quantum mechanical interference of the Bloch states and this
phenomenon has been well studied as the Anderson localization.
According to the scaling theory, all states localize in one- 
or two-dimensional systems without interaction, 
no matter how weak the randomness is.
Interaction between electrons also changes states of electrons drastically.
Strong Coulomb interaction can lead a metallic system to an
insulator when the filling factor is rational and
attractive interaction may cause superconductivity.
Although we have well established understanding
of the localization of non-interacting electrons,
study on the correlated electrons with randomness is
still unsatisfactory\cite{rf:lee}. It is mainly due to lack of
reliable numerical techniques for the correlated electrons with randomness.

Today, for the one-dimensional correlated systems without randomness,
a lot of numerical results are collected by the
Density Matrix Renormalization Group (DMRG)\cite{rf:1} method
and consistent understanding with analytical predictions
has been achieved\cite{rf:d}.
Recently, the DMRG was extended to random systems
 and it has been applied to random spin chains\cite{rf:3}.
It brings us new possibilities to investigate correlation effects on the
random system.
In this paper, we concentrate on dielectric responses of
one-dimentional electronic systems with both randomness and interaction.
Applying a finite electric field $E$, we calculate a polarization $P$
as a derivative of ground state energy
by the field strength $E$. Linear susceptibility $\chi$ is also
obtained numerically as a derivative of the polarization by the electric field.
The polarization $P$ reflects deviation of mass center of electrons
by the field. 
The dielectric response is a fundamental quantity 
to characterize whether the system is a metal or an insulator.
It is suitably defined in open boundary conditions,
while Resta and Sorella defined polarization in periodic system recently.\cite{rf:c1}
Hence the DMRG is naturally applied to investigating the dielectric response.
Using the conventional DMRG, Aebischer {\it et al.} investigated the dielectric responce of Hubbard model with 
next nearest neighbor hopping by
and confirmed the metal-insulator transition of the system.\cite{rf:4}
We apply the DMRG for nonuniform system by Hida\cite{rf:3} to random fermionic systems.

We treat an interacting one-dimensional spinless fermion model in random
potential. 
The Hamiltonian of the $L$ site system is given as
\begin{equation}
H=-t\sum_{i=1}^{L-1} (c_i^\dagger c_{i+1} + h.c.) + V
\sum_{i=1}^{L-1} n_in_{i+1} + \sum_{i=1}^L \epsilon_i n_i,
\end{equation}
where $n_i = c_i^\dagger c_i$ and $\epsilon_i$ is a random potential,
which distributes over an interval $[-W/2,W/2]$ uniformly.
We set $t = 1$ and consider the half-filling case
and impose an open boundary condition.
In the absence of disorder, the system is metallic for $-2<V<2$.
For
half-filling case, at $V=2$ the system undergoes a metal-insulator
transition and the system has a finite charge gap for $V > 2$. In an
attractive interaction region at $V=-2$, the system becomes unstable
due to phase separation.
When the randomness is present, the system is always insulating due to
the Anderson localization without interaction. Then the interplay between the
randomness and interaction can be interesting\cite{rf:ad1,rf:ad2} and possible metallic
phase is expected for some range of negative $V$.
Chui and Bray, and Apel treated the effect of interplay between randomness and
interaction on Tomonaga model.\cite{rf:8,rf:7} The former authors determined critical
value of the interaction constant of localize-delocalize transition by
analysis of the density autocorrelation function. The
latter considered the dynamic conductivity.
The spin-dependent backward scattering was also treated by Chui and 
Bray\cite{rf:9a}, Apel and Rice\cite{rf:9b}, and Suzumura and Fukuyama.\cite{rf:9}
Giamarchi and Schulz took into account the renormalization of
interaction by the disorder and obtained a phase diagram.\cite{rf:6}
Runge and Zymanyi, and Bouzerar and Poilblanc estimated the size of
delocalized region in random interacting spinless fermion model
by the exact diagonalization.\cite{rf:a,rf:2}
Schmitteckert {\it et al.} used the DMRG and estimated the size of the
delocalized region from analysis of the phase sensitivity.\cite{rf:5}

In sec. II, we discuss on the dielectric responses of systems.
Sec. III is for results and discussion. 
Sec. IV is a summary.

\section{Dielectric response}

We focus on dielectric responses of the one dimensional system.
In order to observe the dielectric response directly, we apply the electric field $E$
to
the system. As a second quantized form of the potential, $-Ex$,
the coupling term 
\begin{equation}
 H_E = - E \sum_{i = 1}^L\left(i - \frac{L+1}{2}\right) en_i,
\end{equation}
is added to the Hamiltonian, where $e$ is the charge of the electron and we set $e = 1$ in this paper.  
Then the full
Hamiltonian of the system is given by $H_T = H + H_E$.
As a function of $E$, the polarization $P$ of
 the system is defined as
\begin{eqnarray}
 P =- \frac{1}{L} \frac{ \partial E_{0}}{\partial E }
&=&- \frac{1}{L} \left\langle \frac{\partial H_E}{\partial E}
\right\rangle _E \\ \nonumber
&=& \frac{1}{L} \sum_{i = 1}^L \left(i-\frac{L+1}{2}\right) \langle n_i  \rangle_E,
\end{eqnarray}
where $E_0$ is the ground state energy and $\langle n_i  \rangle _E$
represents the ground state expectation value of $n_i$\cite{rf:4}.
Here we used the Feynman's theorem to derive the expression.

In the Mott insulator, for a finite value of the electric field 
whose potential energy is
comparable with the Mott gap, we expect a collapse of the charge
gap due to the interaction.
Then one can expect a gap closing which is a collapse of the
Mott gap at $E_{\rm g} \sim EL$ where $E_g$ is the Mott gap.
In the Anderson insulator, reconstruction of the charge occurs by
transferring electrons above the tunneling barrier. 
The critical field strength $E_{\rm c}$ is estimated as $E_{\rm c}
\sim E_{\rm g}^{\rm A} / L$ where $E_{\rm g}^{\rm A}$ is an energy difference between
the highest occupied and the lowest empty one particle states.
$E_{\rm g}^{\rm A}$ is of the order of $1/L$ on the average.
In each case, we can obtain information on the charge degree of
freedom above its ground state.

As for a linear response regime, $E \to 0$, we calculate
zero-field dielectric susceptibility as
\begin{equation}
\chi = \left. \frac{\partial P}{ \partial E} \right|_{E = 0}
= - \left. \frac{1}{L} \frac{\partial^2 E_0}{\partial E^2} \right|_{E=0}.
\end{equation}

From the susceptibility $\chi$, we directly obtain information whether
the ground state is metallic or an insulator.
In the thermodynamic limit, $\chi$ is diverging if the system is metallic, but
converging to a finite value if it is an insulator.
Indeed $\chi \sim L^2$ is expected by the perturbation calculation for the pure
non-interacting system($W=0,V=0$). It is also confirmed numerically for
the pure interacting systems($W=0,-2<V<0$) by the DMRG(See later).

In order to calculate the charge distribution and the ground state
energy, we use the exact diagonalization for $V=0$, and the DMRG
for the finite $V$.
In the application of the DMRG, we use the extended infinite-size
algorithm by Hida\cite{rf:3}, which enables us to treat
non-uniform lattice models also.

\section{Results}
\subsection{Response of the two different insulators}
\subsubsection{Mott insulator}

\begin{center}
  \begin{figure}[hbp]
\hspace*{0cm}
   \includegraphics[scale=0.6]{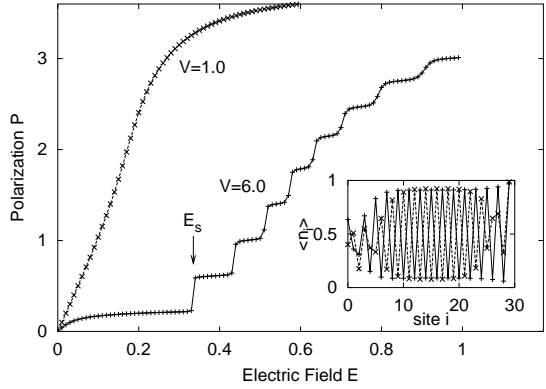}
    \caption{Polarization $P$ as a function of the applied electric field
$E$. (a) For the pure system($W = 0$). 
The Luttinger liquid regime ($V = 1.0$) and
the Mott insulator regime ($V = 6.0$). The system size is $L = 30$. (b)
The charge distribution at each site,
solid line :$E=E_{\rm s} - \Delta E$, broken line :$E=E_{\rm s} + \Delta E$ where
    $E_{\rm s}$ is the value of the electric field where the stepwise behavior is
    observed.}
    \label{fig2}
\end{figure}
\end{center}

We calculate the polarization $P$ in the presence of interaction for
the pure systems by the DMRG.
For $V=1.0$, where the ground state of the system is the
Luttinger liquid.
Then the $P-E$ curve is smooth as shown in Fig.\ \ref{fig2}\cite{rf:loc}.
However for $V=6.0$, where the system is in a Mott insulator phase ,
the $P-E$ curve exhibits a stepwise behavior.
Since the polarization corresponds to the mass
center of the electrons, these steps represent a discontinuous change
 of the charge configuration.
These steps actually come from the generation of kink-pairs.
In the inset of Fig.\ \ref{fig2}, the charge distribution at $E=E_{\rm s} \pm \Delta
E$ are shown, where $E_{\rm s}$ is the smallest value of the electric field
at which the kinks are generated.
At the step, the kink-pair is generated
and the electrons between the kinks are shifted by one site to the right which
compensates for the collapse of the Mott gap.
When the kinks are separated by length $l$, 
the number of electrons between kinks is $l/2$. Then the energy gain
of the length $l$ kinks is
estimated as $\sim El/2$.
Then the first step is due to the kinks with longest length $L$.

\subsubsection{Anderson insulator}

\begin{center}
\begin{figure}
   \includegraphics[scale=0.6]{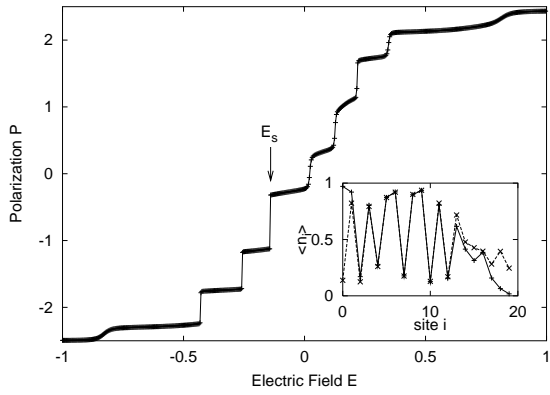}
    \caption{The polarization $P$ as a function of the applied electric field
$E$, non-interacting system($V = 0$,$W=5$,$L=20$). Inset: the charge
   distribution at $E=E_{\rm s} + \Delta E$(broken line) and $E=E_{\rm
    s} -
   \Delta E$(solid line).
}
    \label{fig2b}
\end{figure}
\end{center}

We also calculated the polarization $P$ for the Anderson insulator.
Fig. \ref{fig2b} is the $P-E$ curve for the randomness strength $W=5$ without interaction.
In the presence of randomness, the $P-E$ curve 
exhibits a stepwise behavior also
,which is caused by crossing of one particle localized states.
Matrix elements of the Hamiltonian between arbitrary localized states
are exponentially small as a function of the distance.
Therefore energy repulsion between the highest occupied and the
lowest unoccupied states at both ends of the system is practically
negligible.
Then charge reconstruction of the ground state occurs
which corresponds to the electron tunneling between the localized states.
The inset of Fig.\ \ref{fig2b} is the charge distributions
at $E = E_{\rm s} \pm \Delta E$ where $E_{\rm s}$
is the field strength of the step.
The charge distribution near the center of the lattice is unchanged 
when $E$ crosses the step, but that near the edges is modified. 

\subsection{Linear Response Regime}
\subsubsection{Susceptibility of non-interacting systems}
\begin{center}
\begin{figure}
   \includegraphics[width=0.45\linewidth]{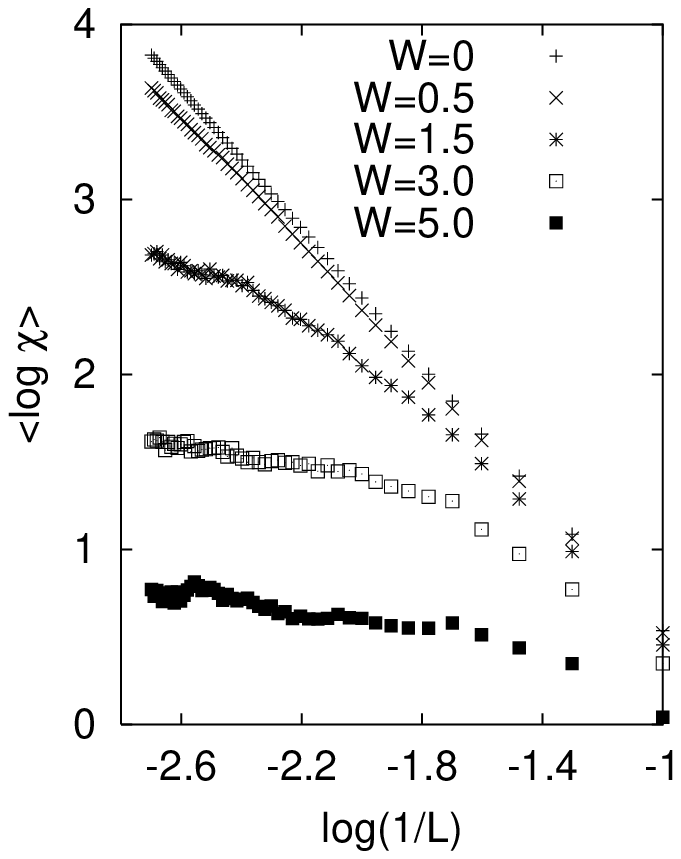}
   \includegraphics[width=0.45\linewidth]{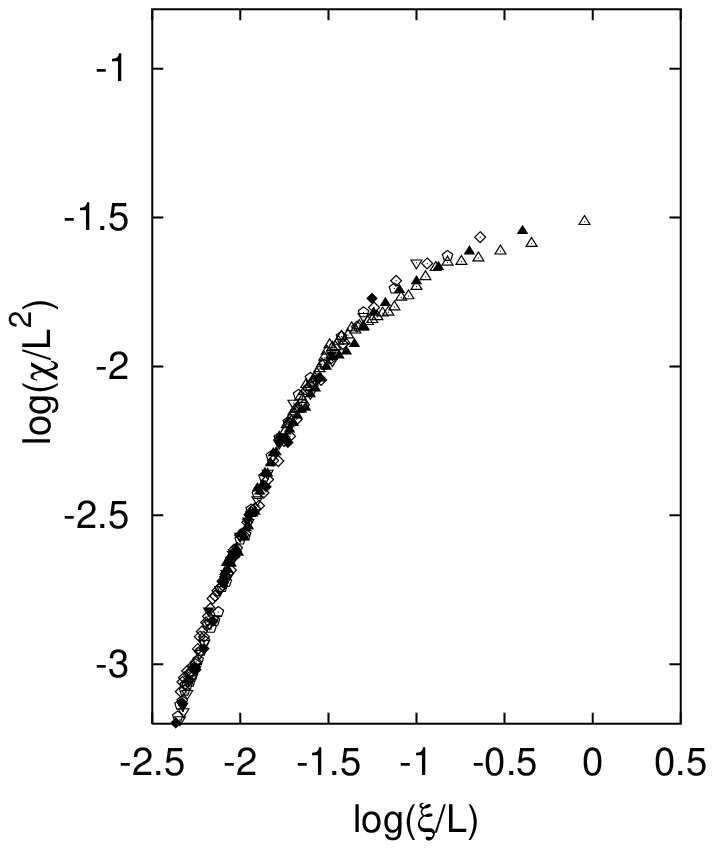}
    \caption{Left: Average $\log\chi$ as a function of $\log1/L$.
    The system sizes are between 10 and 500. $V=0$. $W =$ 0, 0.5,
1.5, 3.0 and 5.0. We averaged over 500 realizations of the disorder potentials.
Right: The scaling function $g$ for $V=0$ is shown.  }
\label{fig3}
\end{figure}
\end{center}

In this section we focus on the linear response region($E \ll t/L$).
We calculate the susceptibility $\chi$ as a function of $L$
by numerically differentiating the ground state energy.
At first, we consider the non-interacting case($V=0$).
In order to calculate $\chi$, we apply small electric field
$\pm \Delta E$. $L \Delta E \sim 
10^{-3}$. 
We need to avoid the occurence of the step in $\pm \Delta E$.
When the step occurs near $E=0$ accidentally, we do not use the data.
The number of occurence of such step is typically $1 \sim 2$ in 100 samples
when $W$ is large($W \sim 5$).
The calculations are always carried out in the localized region except
$W=0$, then we expect $\chi \sim e^{-\xi /L}$ when the system size is
sufficiently large.
Therefore we take an average of $\log \chi $ as is shown in Fig.\ \ref{fig3}.
We averaged over 500 realizations of the disorder potentials.
For $W=0$, $\chi$ increases as $\chi \sim L^2$ and
$\chi$ seems to diverge. This implies the system is metallic. On the
other hand, for $W=3$, $\chi$ is convergent to a finite value
which is consistent with the insulating ground state.
 For intermediate $W$,
saturation of $\chi$ is not clearly observed up to $L = 500$ and
only deviation from $\chi \sim L^2$ is observed.
This is because the localization
length $\xi$ is larger than the system length we used.

Therefore we need a finite-size scaling analysis to determine localization
length $\xi$.
We perform the finite-size scaling analysis by assuming
\begin{equation}
\frac{\chi (L)}{L^2} = g\left( \frac{\xi(W)}{L} \right),
\end{equation}
where $\xi(W)$ is a localization length and $g(x)$ is a scaling
function.
If $L$ is much larger than $\xi$, $\chi$ becomes constant.
Therefore the scaling function $g$ behaves as $g(x)\sim x^2$ for $x \to 0$.
On the other hand, when $L$ is much smaller than $\xi$ $(L \ll \xi )$, the
 wavefunction spread over entire system and 
the system looks like metallic. Then,
$\chi \sim L^2$ is expected and
the scaling function $g$ behaves as $g(x) \sim const.$ for $x \to \infty$.
The scaling function $g$ we obtained for $V=0$ is shown in Fig. \ref{fig3}.
It shows the above-mentioned one parameter scaling hypothesis works quite well
in the model.

\subsubsection{Metal-insulator transition in attractive $V$}
\begin{center}
\begin{figure}
   \includegraphics[scale=0.6]{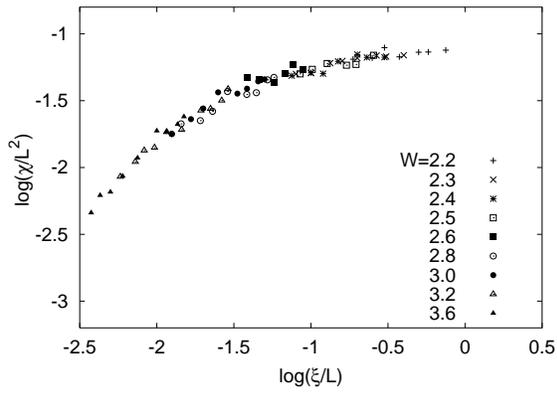}
    \caption{Scaling plots of $\chi / L^2$ vs $ \xi / L$ in a log-log
scale.
$V = -1.4$.  }
\label{fig4}
\end{figure}
\end{center}

Using the random DMRG, we study the systems with attractive electron-electron
interaction.
In our implementation of the DMRG, three or four finite lattice sweeps are
performed to get the convergence of the ground state energy and
the number of retained
states for each block is 60-100 to keep the truncation error to be less
than $10^{-9}$.
The susceptibility $\chi$ is calculated for various values of $V$ and $W$.
The calculations are carried out up to the system size $L=100$. 
We take an average over 128 realizations of the disorder potentials
for $V=-1.4$ and over 64 realizations of the disorder potentials for
$V=-0.5,-0.8,-1.0,-1.1,-1.2,-1.3,-1.5,-1.6$ and $-1.8$.
The steps in $P-E$ curve become smoother when interaction is introduced.
When $\log \chi$ is larger than $5$, we consider this as an influence of
the step and we do not use the data.
Then we perform the finite-size scaling analysis similarly to the
noninteracting case assuming the same one parameter scaling hypothesis.
In order to avoid finite size effect, we use the
data for $L \ge 20$ to determine the localization length $\xi$.
The obtained scaling function $g$ for $V=-1.4$ is shown in Fig.\ \ref{fig4}.
The scaling hypothesis for the interacting case also seems to work
well in the present model.
For $V = -1.8$ the $L$ dependence of $\chi$ deviates from $\chi \sim
L^2$ in $L <
50$ even for $W = 0$.
We could only treat systems $V \ge -1.6$ due to the finite size effects.

\begin{center}
\begin{figure}
   \includegraphics[scale=0.6]{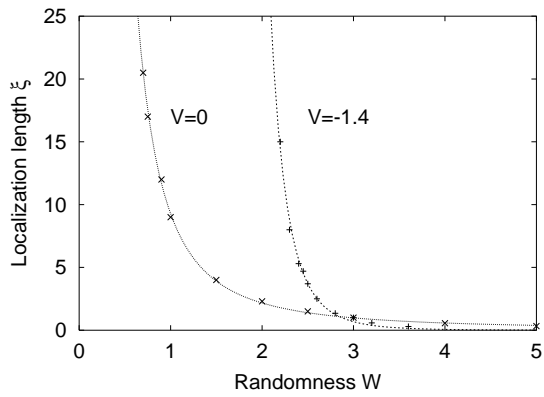}
    \caption{$\xi$ vs $W$. $V=0$ and $-1.4$. $\xi$ is scaled by the value at $W=3$.
The dashed lines show fitting curves by eq (\ref{fit}).
}
\label{fig5}
\end{figure}
\end{center}

In Fig.\ \ref{fig5}, $W$ dependence of the localization length $\xi$ is
shown for various value of $V$.
Note that the localization length $\xi$ is normalized by $\xi (W=3)$.
For $V=0$, it seems that $\xi$ is divergent at $W=0$, but
for $V=-1.4$, $\xi$ increases rapidly with decreasing $W$ and
diverge around $W \sim 1$.

In order to determine the critical disorder strength $W_{\rm c}$,
we fit the localization length $\xi$ as
\begin{equation}
\xi(W) = (A+B(W-W_{\rm c}))(W-W_{\rm c})^\beta , \label{fit}
\end{equation}
where $A,B$ and $\beta$ are fitting parameters.
For $V=0$, we obtained $W_{\rm c}=0.02 \pm 0.09$ and $\beta = -2.1 \pm 0.3$,
which is consistent with expected value $W_{\rm c}=0$ and $\beta =-2$.

\begin{center}
\begin{figure}
   \includegraphics[scale=0.6]{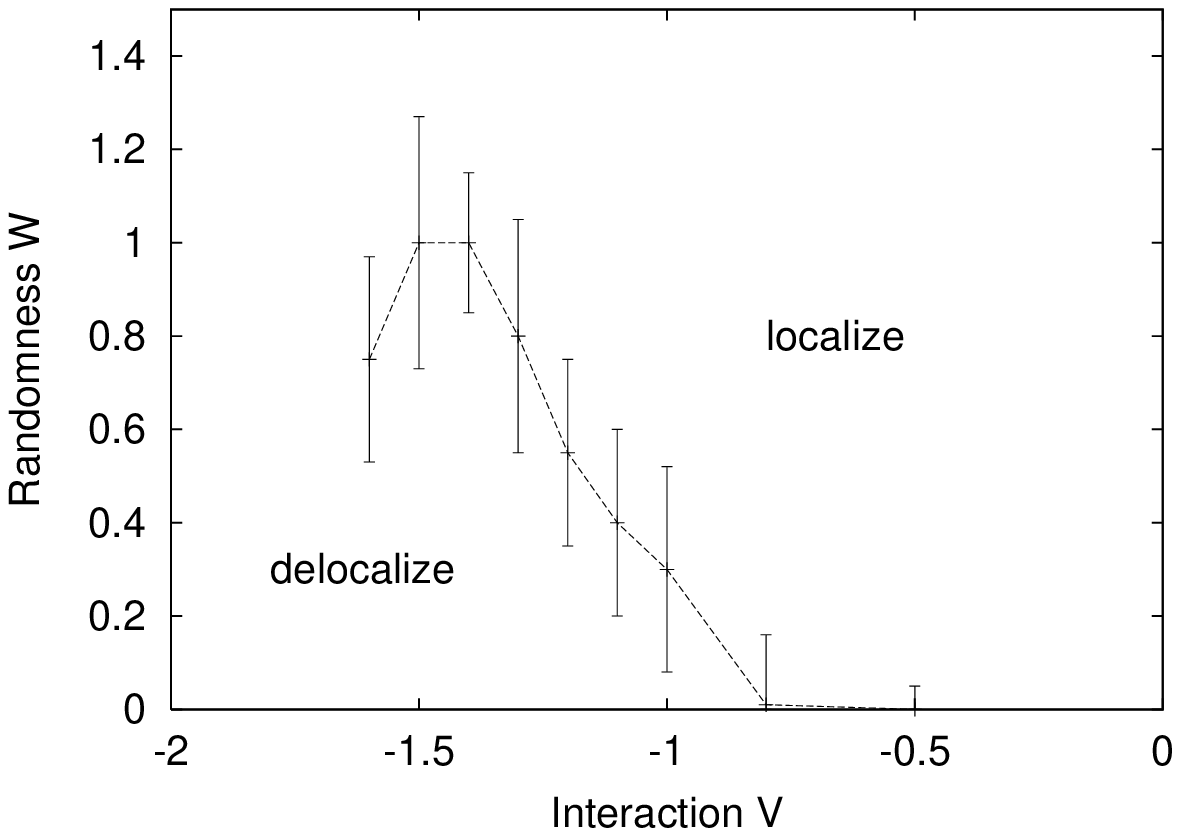}
    \caption{Phase diagram of the half-filled spinless fermion
model. The vertical lines are errorbars.}
\label{fig6}
  \end{figure}
\end{center}

In Fig. \ref{fig6}, the obtained phase diagram is shown.
The vertical lines represent errorbars.
This phase diagram is consistent with the one obtained
from the analysis of the phase sensitibity.\cite{rf:5}

\section{Summary}
In the present paper we have studied the dielectric response of the 
one-dimensional spinless fermion model with interaction and disorder by
using the random DMRG.
At first, we have calculated the polarization of the Mott insulator
and the Anderson insulator.
In the $P-E$ curve we observed stepwise behaviors
both for the Mott insulators and the Anderson insulators. 
From the change of the charge distribution, 
we could understand the stepwise behaviors.
For the Mott insulator the steps come from
the generation of kink-pair and occur at $E \sim 1/L$. 
On the other hand,
the steps for the Anderson insulator represent the crossing of the
energy levels
and the steps occur at $E\sim 1/L^2$.
From the zero field susceptibility, 
we performed the finite-size scaling and determined the localization
length $\xi$.
Also we confirmed the existence of the metallic region 
in attractive interacting regime.

The computation in this work has been done in part using the
facilities of the Supercomputer Center, ISSP, University of Tokyo.
Y. H. was supported in part by a Grant-in-Aid from the Ministry of
Education, Science and Culture of Japan and also by the Kawasaki steel
21st Century Foundation.


\begin{thebibliography}{99}
\bibitem{rf:lee} P. A. Lee and T. V. Ramakrishnan:
Rev. Mod. Phys. {\bf 57} (1985) 287.
\bibitem{rf:1} S. R. White:
Phys. Rev. Lett. {\bf 69} (1992) 2863;
Phys. Rev. {\bf B 48} (1993) 10345.
\bibitem{rf:d} Density-Matrix Renormalization: A New Numerical Method
in Physics, edited by I. Peschel {\it et al}. (Springer-Verlag,
Berlin, 1999).
\bibitem{rf:3} K. Hida:
J. Phys. Soc. Jpn. {\bf 65} (1996) 895.
\bibitem{rf:c1} R. Resta and S. Sorella,
Phys. Rev. Lett. {\bf 82} (1992) 370.  
\bibitem{rf:4} C. Aebischer, D. Baeriswyl and R. M. Noack:
Phys. Rev. Lett. {\bf 86} (2001) 468.
\bibitem{rf:ad1} S. Fujimoto, N. Kawakami:
Phys. Rev. {\bf B 54} (1996) R11018.
\bibitem{rf:ad2} Y. Otsuka and Y. Hatsugai:
J. Phys, Condens. Matter {\bf 12} (2000) 9317.
\bibitem{rf:8} S. T. Chui and J. W. Bray:
Phys. Rev. {\bf B 16} (1977) 1329.
\bibitem{rf:7} W. Apel:
J. Phys. C {\bf 15} (1982) 1973.
\bibitem{rf:9a} S. T. Chui and J. W. Bray:
Phys. Rev. {\bf B 19} (1979) 4020.
\bibitem{rf:9b} W. Apel and T. M. Rice:
Phys. Rev. {\bf B 26} (1982) 7063.
\bibitem{rf:9} Y. Suzumura and H. Fukuyama:
J. Phys. Soc. Jpn. {\bf 52} (1983) 2870.
\bibitem{rf:6} T. Giamarchi and H. J. Schulz:
Phys. Rev. {\bf B 37} (1988) 325.
\bibitem{rf:a} K. J. Runge and G. T. Zimanyi:
Phys. Rev. {\bf B 49} (1994) 15212.
\bibitem{rf:2} G. Bouzerar and D. Poilblanc:
J. Phys. I (France) {\bf 4} (1994) 1699.
\bibitem{rf:5} P. Schmitteckert {\it et al.}:
 Phys. Rev. Lett. {\bf 80} (1998) 560.
\bibitem{rf:loc} M. Kishi and Y. Hatsugai:
J. Phys. Soc. Jpn. {\bf 72} (2003) Suppl. A pp.147.
\end{thebibliography}
\end{document}